\documentclass[prd,amsmath,amssymb,showpacs,superscriptaddress,nofootinbib,twocolumn]{revtex4-1}
\usepackage{graphicx}
\usepackage{epstopdf}
\usepackage[mathlines]{lineno}
\usepackage{dcolumn}
\usepackage{color}
\usepackage{bm}
\usepackage{overpic}
\usepackage{rotating}
\usepackage{times}
\usepackage{multirow}
\usepackage{float}
\usepackage{mathrsfs}
\usepackage{booktabs}
\usepackage[mathlines]{lineno}
\usepackage{subfiles}
\usepackage{orcidlink}

\newcommand{\pipi}{\pi^+\pi^-}

\newcommand{\EE}{e^+e^-}
\newcommand{\pipipi}{\pi^+ \pi^- \pi^0}

\newcommand{\jpsi}{J/\psi}

\newcommand{\BBbar}{B\bar{B}}

\newcommand{\X}{X(3872)}

\newcommand{\B}{\mathcal{B}}
\newcommand{\DDbar}{D^{0}\bar{D}^{0}}
\newcommand{\mbc}{M_{\rm bc}}
\newcommand{\gev}{{\rm GeV}/c^2}
\DeclareMathOperator{\cov}{cov}

\begin{document}

\DeclareGraphicsExtensions{.eps,.png,.ps}

\vspace{0.2cm}

\begin{abstract}

We present a search for the decay $X(3872) \to \pi^+\pi^-\pi^0$ in the
$(772\pm11)\times10^6$ $\Upsilon(4S)\to\BBbar$ data sample collected at the Belle detector,
where the $\X$ is produced in $B^{\pm}\to K^{\pm}X(3872)$ and $B^{0}\to K_{S}^0 X(3872)$ decays.
We do not observe a signal, and set 90\% credible upper limits for two different models of the decay processes: if the decay products are distributed uniformly in phase space, 
$\B(X(3872) \to \pi^+\pi^-\pi^0) < 1.3\%$;
if $M(\pipi)$ is concentrated near the mass of the $\DDbar$ pair in the process
$\X\to D^0\bar{D}^{*0}+c.c.\to\DDbar\pi^0\to\pipi\pi^0$,
$\B(X(3872) \to \pi^+\pi^-\pi^0) < 1.2\times10^{-3}$.

\end{abstract}
\hyphenpenalty=10000
\tolerance=1000
\normalsize
\parskip=5pt plus 1pt minus 1pt
\title{\boldmath Search for $X(3872)\to\pi^+\pi^-\pi^0$ at Belle}

\noaffiliation
 \author{J.~H.~Yin\,\orcidlink{0000-0002-1479-9349}} 
  \author{E.~Won\,\orcidlink{0000-0002-4245-7442}} 
  \author{I.~Adachi\,\orcidlink{0000-0003-2287-0173}} 
  \author{H.~Aihara\,\orcidlink{0000-0002-1907-5964}} 
  \author{S.~Al~Said\,\orcidlink{0000-0002-4895-3869}} 
  \author{D.~M.~Asner\,\orcidlink{0000-0002-1586-5790}} 
  \author{H.~Atmacan\,\orcidlink{0000-0003-2435-501X}} 
  \author{V.~Aulchenko\,\orcidlink{0000-0002-5394-4406}} 
  \author{T.~Aushev\,\orcidlink{0000-0002-6347-7055}} 
  \author{R.~Ayad\,\orcidlink{0000-0003-3466-9290}} 
% ...
\collaboration{The Belle Collaboration}

\maketitle

\section{Introduction}

The state $\X$, also known as the $\chi_{c1}(3872)$, was first observed in 2003 by the Belle Collaboration~\cite{FirstObservationOfX} in the process $B\to K\X,~\X\to\pipi\jpsi$.
The nature of this state remains controversial. For example,
the mass of the $\X$ is very close to the $D^0\bar{D}^{*0}$ threshold~\cite{PDG}, which suggests it could be a $D^0\bar{D}^{*0}$ molecule~\cite{molecu}, but the large production rate in $pp$ and $p\bar{p}$ collision experiments suggests it should have a charmonium core~\cite{pp1,pp2,pp3,pp4}.

Since its discovery, there have been many experimental measurements of the properties of the $\X$ state, including the mass, width, and quantum numbers~\cite{JPC_X3872_1,JPC_X3872_2}.
The recent BaBar measurement of the absolute branching fraction of $B\to K\X$~\cite{Lees:2019xea} makes it possible to obtain the absolute branching fractions of $\X$ decays.
According to a global fit to the branching fraction data~\cite{Li:2019kpj},
the dominant $\X \to D^0\bar{D}^{*0}+c.c.$ decays account for $52^{+25}_{-14}$\% of the decay width and $32^{+18}_{-32}$\% remains unmeasured.

Study of additional $\X$ decay modes could help us understand the components within the $\X$ wave function.
All known $\X$ decays contain open charm or charmonium mesons in the final state, so searches for decays to final states without heavy flavour are of great interest.
Models in which the $\X$ is a charmonium state predict a significant branching fraction for $\X\to gg\to$ light hadrons.
The authors of Ref.~\cite{xdecay_pred} predict that the branching fraction of $\X\to\pipipi$ could be at the level $10^{-3}\sim10^{-4}$ due to the process $\X\to D^0\bar{D}^{*0}\to\DDbar\pi^0\to\pipi\pi^0$~\cite{CCNote}, where the two charged pions come from the rescattering and annihilation of the $\DDbar$ pair.
In this case the main contribution comes from the production of the $\pipi$ pair in a narrow interval of invariant mass $M(\pipi)$ near the mass of the $\DDbar$ pair.

In this paper, we report the {results of a search} for $\X\to\pipipi$ based on $(772\pm11)\times10^6$ $\BBbar$ events collected with the Belle detector,
where the $\X$ is produced in $B^{+}\to K^{+}\X$ and $B^{0}\to K_S^0 \X$ decays.

\section{Belle detector and data samples}
This measurement is based on the full $\Upsilon(4S)$ data sample collected with the Belle detector at the KEKB asymmetric-energy $\EE$ collider~\cite{kekb}.
The Belle detector~\cite{belle} is a large-solid-angle magnetic
spectrometer that consists of a silicon vertex detector
(SVD), a 50-layer central drift chamber (CDC), an array
of aerogel threshold Cherenkov counters (ACC), a barrel-like arrangement of time-of-flight scintillation counters (TOF), and an
electromagnetic calorimeter (ECL) consisting of CsI(Tl)
crystals. All these detector components are located inside a superconducting solenoid coil that provides a 1.5 T
magnetic field. An iron flux-return located outside of the
coil is instrumented with resistive plate chambers to detect $K^0_L$ mesons and to identify muons.
Two inner detector configurations were used: a 2.0 cm beam-pipe and
a 3-layer SVD (SVD1) were used for the first sample of
$152 \times 10^6$ $\BBbar$ pairs, while a 1.5 cm beam-pipe, a 4-layer
SVD (SVD2), and small cells in the inner layers of the
CDC were used to record the remaining $620 \times 10^6$ $\BBbar$
pairs~\cite{svd2}.

The {\sc evtgen}~\cite{evtgen} generator is used
to produce simulated Monte Carlo (MC) events.
The parameters of the $\X$ state in the MC production are taken from Ref.~\cite{PDG}.
The simulation of the detector as well as the response of the particles in the detector are handled with {\sc geant3}~\cite{geant3}.
Two kinds of signal MC events are generated to model the $\X\to\pipipi$ decay.
In the first sample (``case I''), $\X$ decays to three pions are distributed uniformly in phase space.
In the second sample (``case II''), the $\pipi$ invariant mass peaks close to the $\DDbar$ threshold~\cite{xdecay_pred}.
This is implemented in the simulation using an intermediate, dummy, Breit-Wigner resonance with 
a mass of $3729.8~{\rm MeV}/c^2$ and a width of $0.2~\rm MeV$, which are estimated from the prediction.
Backgrounds are studied using generic MC samples: 
$\EE\to q \bar{q},~q=u,~d,~s,~c$ continuum events, and $\EE\to\Upsilon(4S)\to\BBbar$ events with subsequent $b\to c$ decays,
corresponding to twice the integrated luminosity of Belle,
and events with $B$ meson decays to charmless final states, corresponding to 25 times the integrated luminosity.
A tool named {\sc topoana}~\cite{topoana} is used to display the MC event types after event selection.

\section{Event selection}

Charged particle tracks are required to have impact parameters perpendicular to and
along the beam direction with respect to the interaction point (IP)
of less than 1.0 and 3.5~cm, respectively.
Tracks are also required to have at least two hits in the SVD.
Kaons and pions are distinguished using likelihoods based on the response
of the individual sub-detectors~\cite{PID}.
Particles with $\mathcal{R}(K)\equiv \frac{\mathcal{L}(K)}{\mathcal{L}(K)+\mathcal{L}(\pi)}>0.6$, corresponding to a selection efficiency of 80.0\% and a misidentification rate of 7.2\%, are identified as kaons,
where $\mathcal{L}$ is the likelihood for the particle to be a kaon or pion.
Particles with $\mathcal{R}(K)<0.4$, corresponding to a selection efficiency of 83.9\% and a misidentification rate of 9.7\%, are identified as pions.
For pion candidates, similar likelihood ratios for electron~\cite{eID} and muon~\cite{muID} hypotheses are required to be less than 0.1 to further suppress lepton-to-pion misidentification backgrounds.

$K^0_S$ candidates are reconstructed by combining two pions of opposite charge, consistent with emerging from a displaced vertex.
Combinatorial background is suppressed using a neural network~\cite{NNpack,NNKs} utilizing 13 input variables: 
the $K^0_{S}$ momentum in the laboratory frame, the distance along the z axis (opposite the $e^+$ beam
direction) between the two track helices at their closest approach, the $K^0_{S}$ flight length in the transverse plane,
the angle between the $K^0_{S}$ momentum and the vector joining the IP to the $K^0_{S}$ decay vertex, the angle between the pion momentum and the laboratory frame direction in the $K^0_{S}$ rest frame, the distances of closest approach in the transverse plane between the IP and the two pion helices, the number of hits in the CDC for each pion track, and the presence or absence of hits in the SVD for each pion track.
The invariant mass of the two pions is required to satisfy $|M(\pipi)-m_{K^0_{S}}|<0.01~{\rm GeV}/c^2$, where $m_{K^0_S}$ is the $K^0_{S}$ mass~\cite{PDG}.
This mass region corresponds to $\pm 3\sigma$ in the mass resolution.
Neutral pion candidates are reconstructed from photons with deposited energy greater than 50 MeV in the barrel region of the ECL (polar angle within the interval $[33^{\circ},~128^{\circ}]$), or greater than 100 MeV in the end-caps.
The invariant mass of the $\pi^0$ candidate is required to be within the interval $[0.115,~0.155]~{\rm GeV}/c^2$, corresponding to an approximately $\pm 3\sigma$ window around the nominal mass.
A mass constrained fit is then performed.

Reconstructed particles are then combined into a $B^+ \to K^+\pipipi$ or $B^0 \to K^0_S \pipipi$ candidate, and fitted to a common
vertex, which is also constrained to lie in the region around the IP.
Candidates passing the vertex fit are retained.
To improve the resolution in $M(\pipipi)$, we constrain the $K\pipipi$ invariant mass to the nominal mass of the $B$ meson;
the resulting $\pipipi$ invariant mass is then used in further analysis.
For other variables, we use the values obtained before the $B$ mass constraint. 

The $\X$ signal region is defined as $M(\pipipi)\in[3.8,~3.95]~\rm GeV/$$c^2$ and $\mbc>5.27~\rm GeV/$$c^2$, where $\mbc\equiv\sqrt{E_{\rm beam}^2/c^4-|\vec{p}_{B}|^{ 2}/c^2}$ is the beam constrained mass; $E_{\rm beam}$ is the beam energy, and
$\vec{p}_{B}$ is the momentum of the reconstructed $B$ meson in the $\EE$ center-of-mass frame.
Separate searches are conducted for $\X\to\pipipi$ decays according to phase space (the case I sample) and for decays according to Ref.~\cite{xdecay_pred} (the case II sample). Up to this point, all selection criteria are common.
In the case II analysis, an extra requirement $M(\pipi)\in[3.7,~3.75]~\rm GeV/$$c^2$ is imposed.

\begin{figure}[htbp]
\begin{center}
    \begin{overpic}[width=0.22\textwidth]{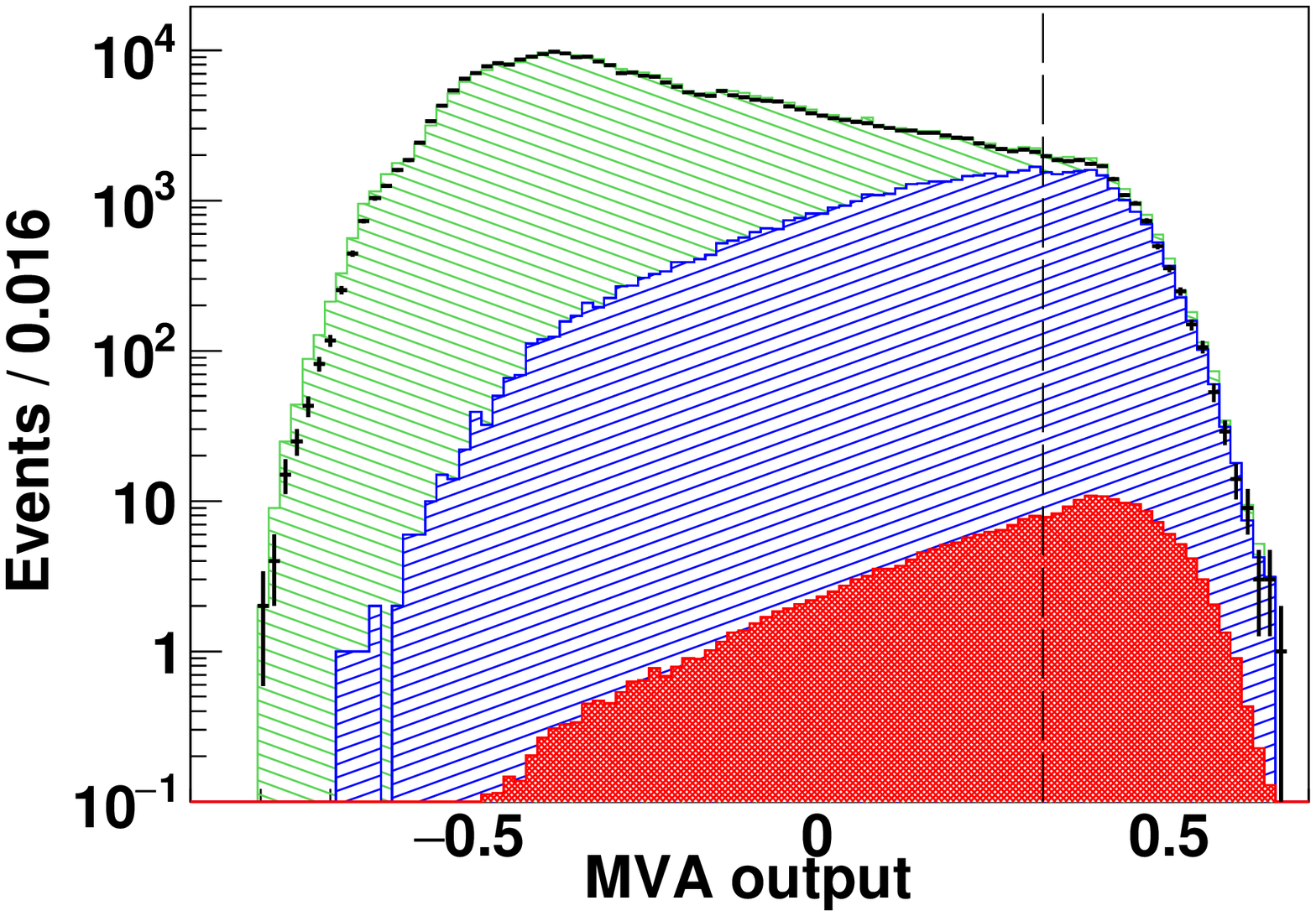}
           \put(20,58){\textbf{(a)}}
    \end{overpic}
    \begin{overpic}[width=0.22\textwidth]{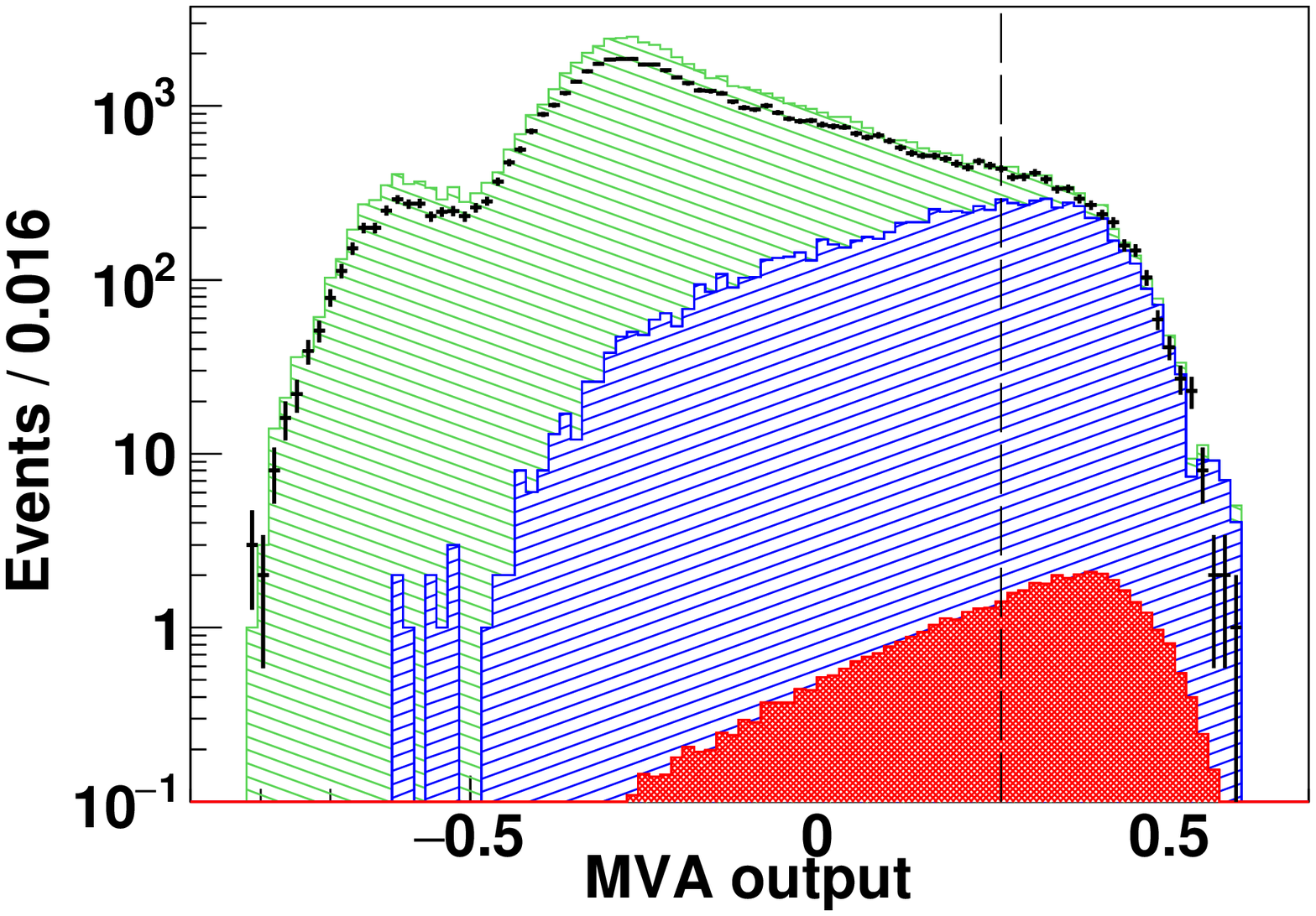}
           \put(20,58){\textbf{(b)}}
    \end{overpic}
    \begin{overpic}[width=0.22\textwidth]{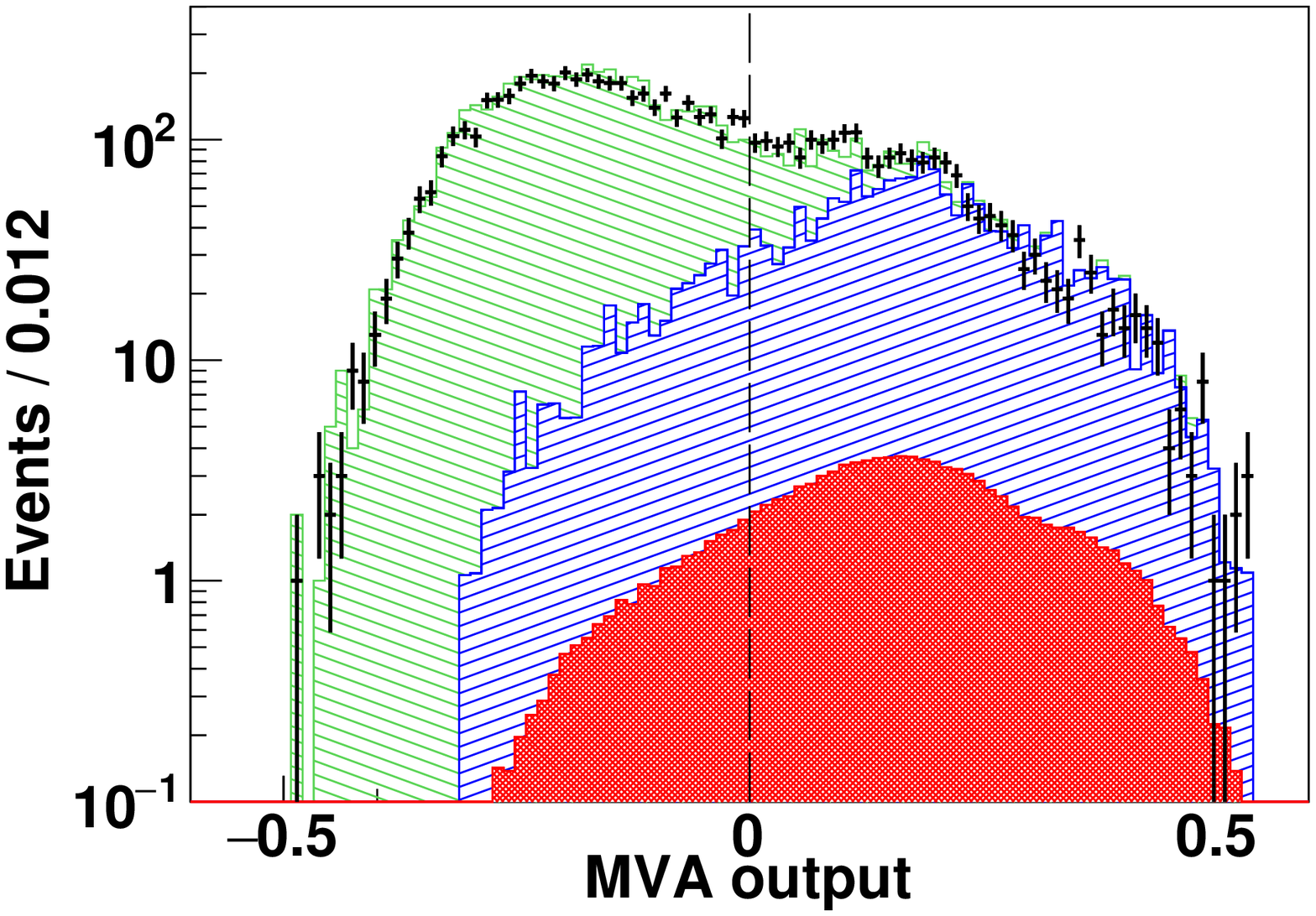}
           \put(20,58){\textbf{(c)}}
    \end{overpic}
    \begin{overpic}[width=0.22\textwidth]{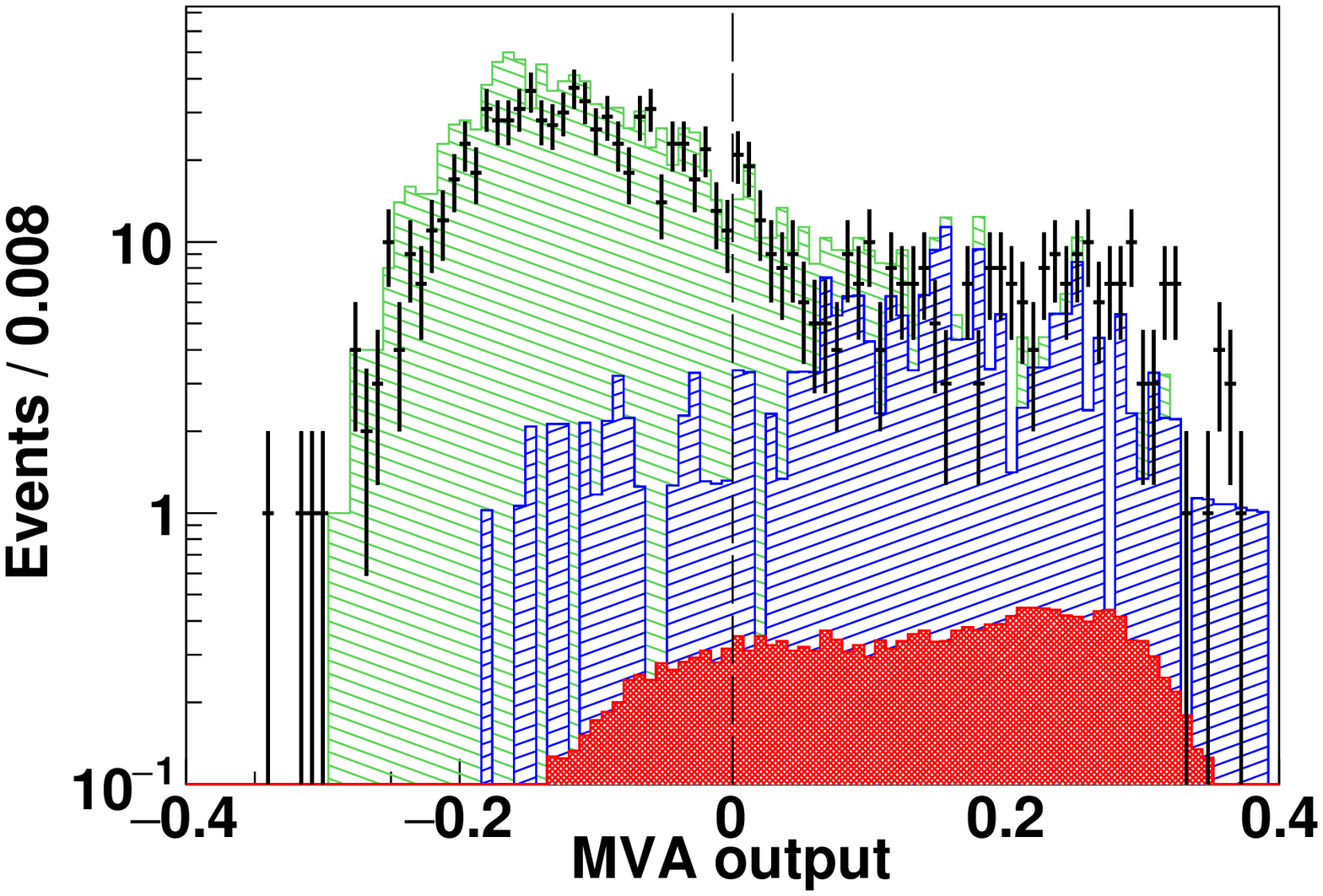}
           \put(20,58){\textbf{(d)}}
    \end{overpic}
    
\end{center}
\caption{Distributions of the MVA discriminator output for (a) the charged mode in case I, (b) the neutral mode in case I, (c) the charged mode in case II, and (d) the neutral mode in case II. Dots with error bars show the experimental data, blue shaded histogram the normalized generic $\BBbar$ MC sample, green shaded histogram the normalized generic continuum MC sample, and red shaded histogram the normalized signal MC sample. Vertical dashed lines represent the requirements applied in the analyses.
 }\label{fig:MVAoutput}
\end{figure}

The largest background arises from continuum production.
We use multivariate analysis (MVA) implemented in {\sc root}~\cite{tmva} to suppress the continuum background with the following variables:
modified Fox-Wolfram moments~\cite{fwmomentum},
the angle between the thrust axis of the $B$ meson candidate and that of the remaining particles in the event,
the angle between the thrust axis of all tracks and the thrust axis of all photons in the event,
the vertex fit quality, including both the vertex fit and the constraint to the IP,
the $B$ meson production angle, the $K$ meson helicity angle, and the invariant mass of the $\pi^0$ meson before the mass constrained fit.
The training and optimization of the MVA are performed with signal and continuum MC samples.
We choose the Boosted Decision Tree as our training method in the MVA.
Distributions of the MVA output are shown in Fig.~\ref{fig:MVAoutput}.
We use the figure of merit $N_{\rm S}/\sqrt{N_{\rm S}+N_{\rm B}}$ to optimize the MVA selection, where $N_{\rm B}$ is the number of background events from the generic MC, and $N_{\rm S}$ is the expected number of signal events estimated according to the predicted branching fraction $1.0\times10^{-3}$.
Both $N_{\rm S}$ and $N_{\rm B}$ are counted in the signal region defined separately in the two different cases.
For case I, an MVA output greater than 0.32 and 0.26 is required for the charged and neutral mode respectively.
For case II, an MVA output greater than 0.0 is required for both charged and neutral modes.
These requirements reject nearly 99\% of the continuum background.

After continuum suppression,
a requirement on the energy difference $\Delta E \equiv E_{\rm beam} - E_{B}$ is applied 
to suppress the background from B meson decays where the wrong combination of particles has been chosen.
Here $E_{B}$ is the energy of the reconstructed $B$ meson.
To suppress $B$ meson decays to the same final state as the signal process, for example, $B\to D \rho$, $B\to K^{*}(892)\rho$, mass window requirements on $M(K^{\pm}\pi^{\mp,0})$ and $M(K^{\pm}\pi^{\mp}\pi^{0})$ are imposed.
These requirements are optimized using a similar figure of merit.
The selection criteria are summarized in Table~\ref{tab:cuts}.
An extremely large mass window on $M(K^{\pm}\pi^{\mp}\pi^0)$ is imposed to veto not only $D^0$ but also $D^{*0}$ and other resonances.
If there are multiple candidates in one event, the candidate with the highest MVA performance is chosen.

\begin{table}[htp]
\caption{Requirements applied to the charged and neutral $B\to K\X$ decay modes in the two analyses.}
\begin{center}
\begin{tabular}{l r r @{\hskip 4pt} r @{\hskip 8pt} r @{\hskip 4pt} r @{\hskip 10pt} l }
\hline
\hline			
					& \multicolumn{3}{c}{case I}					
					& \multicolumn{2}{c}{case II}			&	\\
					& \multicolumn{2}{c}{$B^+$}	
								& \multicolumn{1}{c}{$B^{0}$}	
					& \multicolumn{1}{c}{$B^+$}
								& \multicolumn{1}{c}{$B^{0}$}
											& units	\\
\hline
MVA& $>$					&   $0.32\phantom{0}$	& $0.26\phantom{0}$	
					&   $0.0\phantom{00}$	& $0.0\phantom{00}$	&	\\
$|\Delta E|$ & $<$				&   $0.040$		& $0.045$		
					&   $0.045$		& $0.035$		& ${\rm GeV}$ \\
$|M(K^{\pm}\pi^{\mp}) - m_{D^0}|$ & $>$	&   $0.03\phantom{0}$	&\multicolumn{1}{c}{--~~}
					&   $0.04\phantom{0}$	&\multicolumn{1}{c}{--~~}& ${\rm GeV}/c^2$ \\
$|M(K^{\pm}\pi^{\mp}\pi^{0}) - m_{D^0}|$ & $ >$ &$0.95\phantom{0}$	&\multicolumn{1}{c}{--~~}&
					    $0.55\phantom{0}$	&\multicolumn{1}{c}{--~~}& ${\rm GeV}/c^2$ \\
$|M(K^0_S\pi^{\pm/0}) - m_{D^{\pm/0}}|$ & $>$&\multicolumn{1}{c}{--~~}& $0.020$
					&\multicolumn{1}{c}{--~~}& $0.030$		& ${\rm GeV}/c^2$ \\
$|M(K^0_S\pi^{\pm}\pi^0) - m_{D^{\pm}}|$ & $>$&\multicolumn{1}{c}{--~~}& $0.50\phantom{0}$
					&\multicolumn{1}{c}{--~~}& $0.10\phantom{0}$	& ${\rm GeV}/c^2$ \\
$M(K\pi)$ & $>$     				& 1.0\phantom{00}	& 1.0\phantom{00}
					& 1.0\phantom{00}	& 1.0\phantom{00}	& ${\rm GeV}/c^2$ \\
\hline
\end{tabular}
\end{center}
\label{tab:cuts}
\end{table}%

\section{Data analysis}

After the event selection described above, we find that there is a remaining background, peaking in $\mbc$, from rare charmless $B$ meson decays such as $B\to K^{*}\rho$.
If the MC description of these decays were entirely correct, we would expect the contribution of this background in data to be
$1/25$ $ = 0.04$ times as large as that in the MC sample.
We extract the actual scale factor from data by studying the events in the region $M(\pipipi)\in[3.2,~3.5]~{\rm GeV}/c^2$,
where no charmonium decays to three pions are expected.
Unbinned maximum likelihood fits are performed to the $\mbc$ distributions from the data and MC samples in this region.
The peaking background is described by two Gaussians with a common mean, and other events are described by an ARGUS~\cite{ARGUS} function.
By comparing yields from data and MC samples, the scale factors between the data and rare charmless $B$ meson decay MC samples are extracted, as listed in Table~\ref{tab:Npeak}.

\begin{table}[htp]
\caption{Peaking background yields in the rare charmless $B$ meson decay MC and data samples, and the resulting scale factor, for the charged and neutral modes.}
\begin{center}
\begin{tabular}{c  c  c}
\hline
\hline
 	 	& 	$B^{+}\to K^{\pm}\X$		&	$B^{0}\to K_{S}^0\X$		\\
\hline
MC sample    &		$5082.5\pm83.8$   		&	$2202.5\pm57.6$ 	\\
data sample	&	$\phantom{5}286.0\pm45.2$	&	$\phantom{2}171.2\pm28.3$ 	\\
\hline
scale factor		&		$(5.05\pm0.82)\times10^{-2}$ & $(7.06\pm1.22)\times10^{-2}$ \\
\hline
\end{tabular}
\end{center}
\label{tab:Npeak}
\end{table}%

\begin{figure*}[htbp]
\begin{center}
    \includegraphics[width=0.99\textwidth]{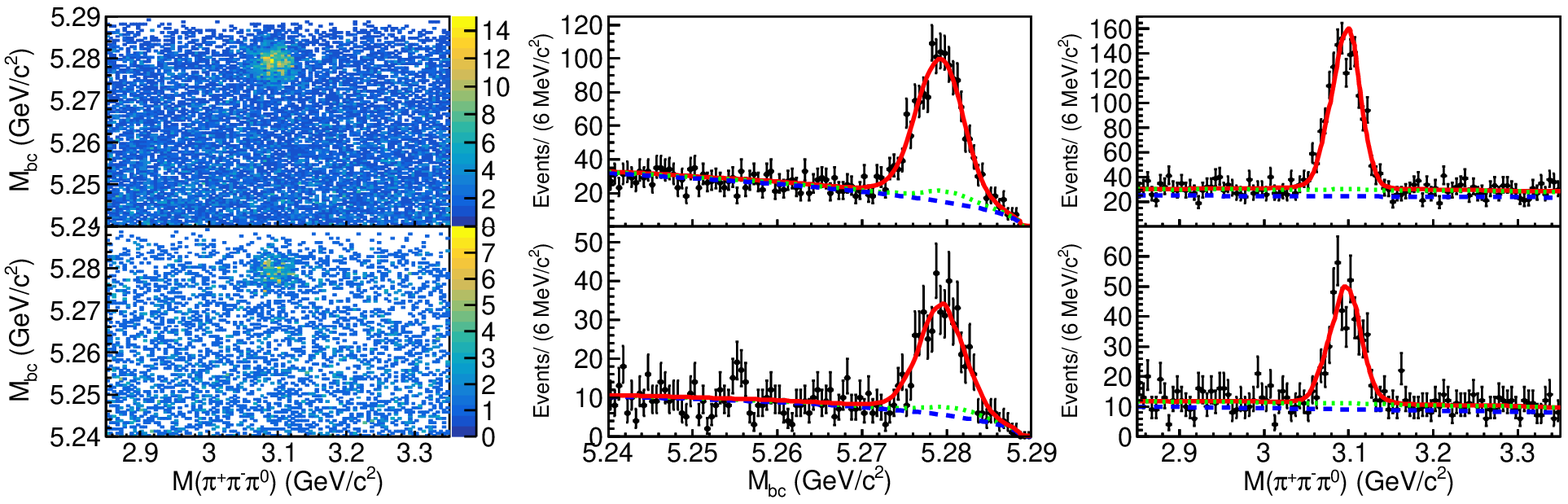}
\end{center}
\caption{Events in the signal region for (top) $B^+ \to K^+ \jpsi$ and (bottom) $B^0 \to K^0_S \jpsi$. Left plots show the ($M(\pi^+\pi^-\pi^0),~\mbc$) distribution, center plots show the
projection on $\mbc$, and right plots show the projection on $M(\pipipi)$. In the
center and right plots, dots with error bars show the experimental data, red curves the fit results, blue dashed curves the combinatorial background, and green dotted curves show the sum of the combinatorial and $B\to K \pi^+\pi^-\pi^0$ backgrounds.}\label{fig:SimFitJpsi}
\end{figure*}

We use the $B\to K\jpsi$ decay to validate our event selection and signal extraction procedures.
The $\mbc$ and $M(\pipipi)$ distributions in the $\jpsi$ signal region, $M(\pipipi)\in[3.05,~3.15]~\rm GeV/$$c^2$ and $\mbc>5.27~\rm GeV/$$c^2$, are shown in Fig.~\ref{fig:SimFitJpsi}.
No correlation between $\mbc$ and $M(\pipipi)$ is found. 
An unbinned two dimensional simultaneous fit is performed to the $(\mbc,M(\pipipi))$ distributions for $B^{+}\to K^+\jpsi$ and $B^{0}\to K_{S}^{0}\jpsi$.
Three components are used in the fit, including $\jpsi$ signal, combinatorial background, and $B\to K \pipipi$ background peaking in $\mbc$.
The signal is described with a MC simulated histogram, smeared in $\mbc$ with a Gaussian representing the discrepancy between data and MC simulation; the width of the Gaussian is allowed to float.
The signal MC simulated histogram is modeled using kernel estimation~\cite{rookeyspdf}.
Signal yields from the charged and neutral modes are converted to branching fractions using the formula:
\begin{equation}
N_{\mathrm obs} = 2\times N_{\BBbar} f \B(B\to K\jpsi) \B(\jpsi\to\pipipi) \epsilon,
\end{equation}
where $N_{\mathrm obs}$ is the observed signal yield, $N_{\BBbar}\equiv772\times10^{6}$ is the number of $\BBbar$ pairs, $f\equiv0.514$ or $0.486$ is the fraction of charged or neutral $\BBbar$ pairs, $\B(B\to K\jpsi)$ is the branching fraction of $B^{+/0}\to K^{+/0}\jpsi$~\cite{PDG}, and $\epsilon$ is the reconstruction efficiency for each mode obtained from the signal MC study; corrections to particle identification (PID) efficiencies, to match those measured in data, are included.
Thus we can extract the branching fraction $\B(\jpsi\to\pipipi)$ directly from the simultaneous fit.
Combinatorial backgrounds are described with an ARGUS function in $M_{\rm bc}$ and a 1st-order-polynomial function in $M(\pipipi)$.
The $B\to K \pipipi$ background is distributed smoothly in $M(\pipipi)$, but peaks in $\mbc$.
In the fit, the shape of the $B\to K \pipipi$ background is extracted from the rare charmless $B$ meson decay MC simulation.
The scaling factors on the normalisation of this background
for the $B^+$ and $B^0$ final states float in the fit, subject to a Gaussian constraint
with mean and uncertainty taken from Table~\ref{tab:Npeak}.
The results of the simultaneous fit are shown in Fig.~\ref{fig:SimFitJpsi}.
The fitted branching fraction, $\B(\jpsi\to\pipipi)=(2.10\pm0.06)\%$,
is consistent with the world average value~\cite{PDG}.

\begin{figure*}[htbp]
\begin{center}
    \includegraphics[width=0.99\textwidth]{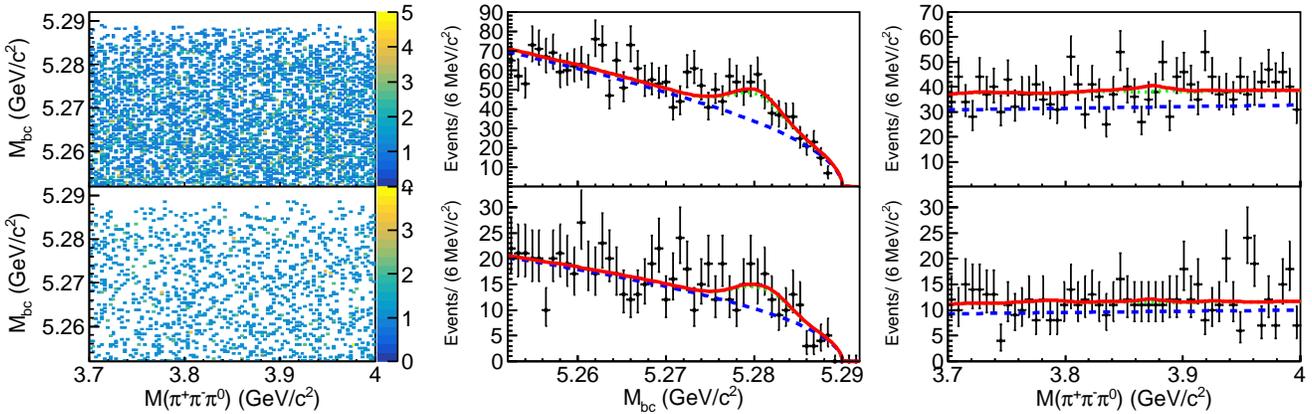}
\end{center}
\caption{Events in the signal region for (top) $B^+ \to K^+ \X$ and (bottom) $B^0 \to K^0_S \X$ in the case I analysis. Left plots show the ($M(\pi^+\pi^-\pi^0),~\mbc$) distribution, center plots show the
projection on $\mbc$, and right plots show the projection on $M(\pipipi)$. In the
center and right plots, dots with error bars show the experimental data, red curves the fit results, blue dashed curves the combinatorial background, and green dotted curves show the sum of the combinatorial and $B\to K \pi^+\pi^-\pi^0$ backgrounds.}\label{fig:SimFitXphsp}
\end{figure*}

The $\mbc$ and $M(\pipipi)$ distributions in the $\X$ signal region are shown in Fig.~\ref{fig:SimFitXphsp} for experimental data in the case I analysis.
We follow the same fitting procedure used for $B \to K J/\psi$.
The parameters of the Gaussian function used to smear the MC signal shape in $\mbc$ are fixed to the results from the $\jpsi$ fit.
No significant signal is found.
The fitted branching fraction $\B(\X\to\pipipi) = (2.6\pm2.8)\times10^{-3}$, corresponding to ($20.3\pm22.0$) $B^{\pm}\to K^{\pm}\X$ and ($4.2\pm4.6$) $B^{0}\to K^{0}\X$ events.
An upper limit for the branching fraction with the systematic uncertainty is estimated using the following method. 
By varying the branching fraction of $X(3872)\to\pipipi$ in the fit, the branching fraction dependent relative likelihood distribution is obtained. 
This likelihood distribution is then convolved with a Gaussian function which models the systematic uncertainty. The upper limit is determined by the value for which the integral of this new PDF is 90\% of its total area.
Estimation of systematic uncertainties is discussed in Section~\ref{sec:systematics}.  
The uncertainty of $\B(B\to K \X)$ is quoted from the global fit~\cite{Li:2019kpj}.
The 90\% credible upper limit is $\B(\X\to\pipipi) < 1.3\%$.

Because of the large systematic uncertainty introduced by the branching fraction of $B\to K\X$,
we also fit the $(\mbc,M(\pipipi))$ distributions of the charged and neutral modes separately to obtain the product of the branching fractions $\B(B\to K\X)\B(\X\to\pipipi)$.
The fit procedure is otherwise the same as that for the simultaneous fit.
The signal yields for the charged and neutral modes are $25.4\pm24.0$ and $-6.2\pm10.6$, respectively,
with corresponding 90\% credible upper limits $N_{\rm up}$ of $61$ and $19$.
Using the formula $\frac{N_{\rm up}}{2N(\BBbar)f\epsilon}$,
the upper limits on the products of branching fractions are calculated to be
$\B(B^+\to K^+\X)\B(\X\to\pipipi)<1.9\times10^{-6}$ and
$\B(B^0\to K^0_S \X)\B(\X\to\pipipi)<1.5\times10^{-6}$.

\begin{figure}[htbp]
\begin{center}
    \includegraphics[width=0.5\textwidth]{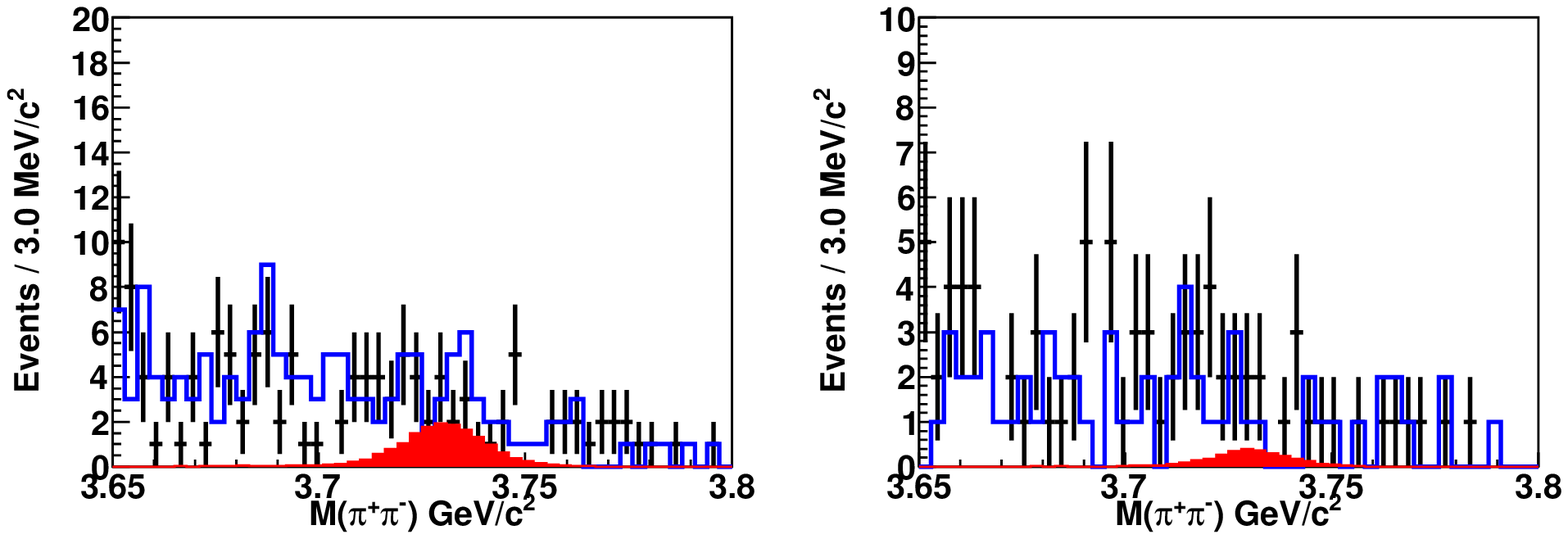}
\end{center}
\caption{$M(\pipi)$ distributions in the $\X$ signal region in the case II analysis. Dots with error bars show the experimental data from the charged (left) and neutral (right) channel. The red shaded histogram shows the signal MC, normalized assuming $\B(\X\to\pipipi)=1\times10^{-3}$. The blue solid line shows the MC background samples, normalized to the same integrated luminosity as the experimental data.
}\label{fig:M23_data}
\end{figure}

\begin{figure*}[htbp]
\begin{center}
    \includegraphics[width=0.99\textwidth]{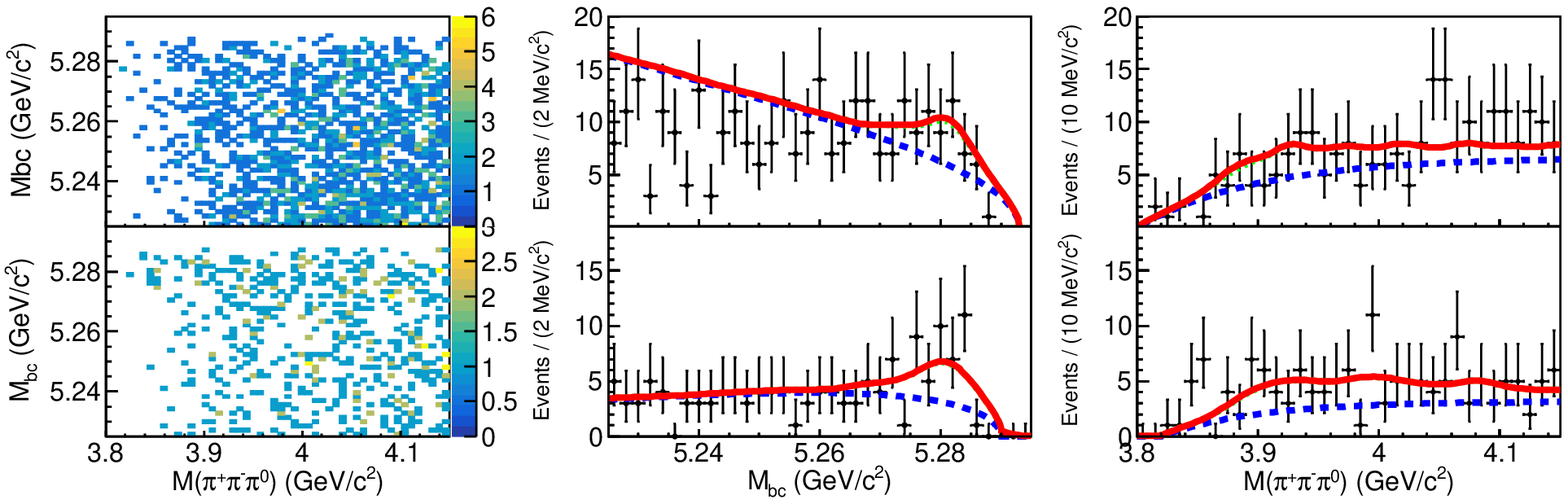}
\end{center}
\caption{Events in the signal region for (top) $B^+ \to K^+ \X$ and (bottom) $B^0 \to K^0_S \X$ in the case II analysis. Left plots show the ($M(\pi^+\pi^-\pi^0),~\mbc$) distribution, center plots show the
projection on $\mbc$, and right plots show the projection on $M(\pipipi)$. In the
center and right plots, dots with error bars show the experimental data, red curves the fit results, blue dashed curves the combinatorial background, and green dotted curves show the sum of the combinatorial and $B\to K \pi^+\pi^-\pi^0$ backgrounds.
}\label{fig:SimFitXpred}
\end{figure*}

For the case II analysis, the $M(\pipi)$ distributions for the charged and neutral modes are shown in Fig.~\ref{fig:M23_data},
for events in the $\X$ signal region in data. No significant enhancement near the $\DDbar$ threshold is found.
After the case II selection criteria, the $\mbc$ and $M(\pipipi)$ distributions for the charged and neutral modes are shown in Fig.~\ref{fig:SimFitXpred}. 
The fit is similar to that used in Case I, except that we use a reversed exponential function ($1-1/e^{p_0 (M(\pipipi)-m_{\rm thres})}$) to describe the combinatorial background shape in $M(\pipipi)$, where $m_{\rm thres}\equiv3.70 + m(\pi^0)~\gev$.
No significant signal is found in this scenario, either.
The scale factors for peaking background are fitted to be $0.058\pm0.008$ and $0.138\pm0.043$ for the charged and neutral mode, respectively.
The fitted branching fraction is $\B(\X\to\pipipi) = (0.9\pm3.1)\times10^{-4}$, corresponding to ($1.5\pm5.4$) $B^{\pm}\to K^{\pm}\X$ and ($0.3\pm1.0$) $B^{0}\to K^{0}\X$ events.
The 90\% credible upper limit, established using the same method as in Case I, is $\B(\X\to\pipipi) < 1.2\times10^{-3}$.
Separate fits to the charged and neutral modes find signal yields of $0.7\pm5.5$ and $5.3\pm5.6$, respectively,
with corresponding 90\% credible upper limits $N_{\mathrm up}$ of $11.2$ and $14.8$.
The upper limits on the products of branching fractions
are calculated to be 
$\B(B^+\to K^+\X)    \B(\X\to\pipipi) < 1.5\times10^{-7}$ and
$\B(B^0\to K^0_S \X) \B(\X\to\pipipi) <1.8\times10^{-7}$.

\section{Systematic uncertainty}
\label{sec:systematics}

Possible sources of systematic uncertainty include tracking, PID, $K_S^{0}$ reconstruction, $\pi^0$ reconstruction, the signal MC generation model, the MVA requirements, signal yields, the number of $\BBbar$ events, and $B\to K\X$ branching fractions.

The difference in tracking efficiency for momenta above $200~\rm MeV/c$ between data and MC is $(-0.13\pm0.30\pm0.10)\%$ per track.
We apply a reconstruction uncertainty of $0.35\%$ per track in our analysis.
According to the updated measurement of PID efficiency using the control sample $D^{*}\to D^0\pi$ and $D^0\to K^-\pi^+$, we assign uncertainties of 1.1\% for each kaon and 0.9\% for each pion.
For $K_S^0$ selection, we take 2.2\% as the systematic uncertainty following Ref.~\cite{kserr}.
For $\pi^0$ selection, the uncertainty is 2.3\% according to a study of the $\tau\to\pi\pi^0\nu_{\tau}$ control sample~\cite{pi0err}.

\begin{table}[htp]
\caption{Systematic uncertainties on the $B(\X\to\pipipi)$ measurement (in units of \%). 
}
\begin{center}
\begin{tabular}{l c c}
\hline
\hline
Source			& $B^+\to K^+\X$			& $B^0\to K_{S}^0\X$			\\
\hline
Tracking		& 1.1					& 0.7					\\
PID			& 2.9					& 1.8					\\
$K_{S}^0$ selection	& 0.0					& 2.2					\\
$\pi^0$ selection 	& 2.3					& 2.3					\\
Signal MC model		& 0.7\makebox[0cm][l]{  }		& 0.7\makebox[0cm][l]{ }		\\
$\mathcal{B}(B \to K X(3872))$	& \!\!31.6\makebox[0cm][l]{ }	& \!\!\!\!$^{+45.4}_{-36.4}$\makebox[0cm][l]{ }\\
Total 		&    \!\!31.8\makebox[0cm][l]{ }	& \!\!\!\!$^{+45.6}_{-36.6}$\makebox[0cm][l]{ }\\
\hline
Number of $\BBbar$		& \multicolumn{2}{c}{1.4}					\\
$B^+ B^-$ Fraction		& \multicolumn{2}{c}{1.2}					\\
Selection criteria   		& \multicolumn{2}{c}{5.0}					\\
\hline
Weighted  total			& \multicolumn{2}{c}{$^{+35.7}_{-34.1}$ \makebox[0cm][l]{}}\\
\hline
\end{tabular}
\end{center}
\label{tab:sumsys}
\end{table}%

\begin{table*}[htp]
\caption{Summary of the measured branching fractions. }
\begin{center}
\begin{tabular}{c c c}
\hline
\hline
channel	                         & 	{case I}					&	{case II}					\\
\hline
$B^{\pm}\to K^{\pm}\X,~\X\to\pipipi$	&		$<1.9\times10^{-6}$		&	$<1.5\times10^{-7}$ 		\\
$B^{0}\to K^0\X,~\X\to\pipipi$		&		$<1.5\times10^{-6}$ 		&	$<1.8\times10^{-7}$    \\
\hline
$\X\to\pipipi$		&		$<1.3\%$					&		$<1.2\times10^{-3}$			\\
\hline
\end{tabular}
\end{center}
\label{tab:sumbranch}
\end{table*}%

In the case II analysis, the angular distribution of the decay of the pseudo intermediate state may also affect the reconstruction efficiency.
MC samples with the helicity angle of the intermediate state following $1+\alpha{\rm cos}\theta$, $\alpha=-1,~0,~1$ have been generated.
The reconstruction efficiencies for these samples are consistent with each other within the statistical uncertainty.
Thus no contribution to the systematic uncertainty is added from this source.
The width of the intermediate state used in our generator may also affect the reconstruction efficiency.
We broaden the lineshape from a width of 0.2 MeV to 1.0 MeV, and find a difference in the reconstruction efficiency of only 0.7\%.

We test our selection criteria listed in Table~\ref{tab:cuts} with the control sample $B\to K\jpsi$, and the extracted branching fraction of $\jpsi\to\pipipi$ is consistent with the world average value~\cite{PDG} within the statistical uncertainty of the fit. We assign this uncertainty as the systematic uncertainty due to the selection criteria.

The systematic uncertainties on the signal yields are due to the signal and background descriptions.
For the signal part, the discrepancy between data and MC simulation is represented with a Gaussian function obtained from the validation sample.
By varying the width of the convolution Gaussian by $\pm 1\sigma$, the upper limits on the signal yields do not change. Thus no systematic uncertainty is assigned.
For the background, we vary the combinatorial background shape descriptions as well as the fit range, and choose the largest upper limit estimation as the most conservative result.

The systematic uncertainty on the number of total $\BBbar$ events is taken as 1.4\%, and the systematic uncertainty on the fraction of charged and neutral $\BBbar$ is taken as $1.2\%$~\cite{PDG}.
The systematic uncertainty on the $B\to K\X$ branching fractions are taken as 31.6\% and $^{+45.4}_{-36.4}$\%~\cite{Li:2019kpj} for the charged mode and neutral modes, respectively.

We summarize the systematic uncertainties in Table~\ref{tab:sumsys}.
The total systematic uncertainty for the joint branching fraction from the charged or neutral mode is obtained by adding the individual
components in quadrature.
The total systematic uncertainty of the $\X\to\pipipi$ branching fraction measurement is calculated with the following formula:
\begin{equation}
    \sigma =\frac{\sqrt{\sum_{i=1}^{2}
    (\mathcal{B}_{i}\epsilon_{i}\sigma_{i})^{2}+2\cov(1,2)}}{\sum_{i=1}^{2}
    \mathcal{B}_{i}\epsilon_{i}},
\end{equation}
where the branching ratio is taken from Ref.~\cite{PDG}; $\epsilon_{i}$
is the reconstruction efficiency and $\sigma_{i}$ is the
systematic uncertainty for each mode, which is obtained by adding each components in quadrature;
$\cov(1,2)=\Pi_{i=1}^{2}\mathcal{B}_{i}\epsilon_{i}\sigma_{i}^{\prime}$
is the covariance systematic uncertainty between the two modes, in which $\sigma^{\prime}$ is the common systematic uncertainty.


\section{Summary}

We have carried out a search for the decay $\X\to\pipipi$ in the 
$(772\pm11)\times10^6$ $\Upsilon(4S)\to\BBbar$ data sample collected at the Belle detector,
in $B\to K\X$ events. No signal is seen.
We set 90\% credible upper limits on the branching fraction in two different models of the decay:
if the decay products are distributed uniformly in phase space, 
$\B(X(3872) \to \pi^+\pi^-\pi^0) < 1.3\%$;
if $M(\pipi)$ is concentrated near the mass of the $\DDbar$ pair in the process
$\X\to D^0\bar{D}^{*0}+c.c.\to\DDbar\pi^0\to\pipi\pi^0$,
$\B(X(3872) \to \pi^+\pi^-\pi^0) < 1.2\times10^{-3}$.
Upper limits on the product branching fractions
$\B(B\to K \X)\B(X\to\pipipi)$ are also set for both charged and neutral $B$ decays, as listed in Table~\ref{tab:sumbranch}.
This measurement may be used to provide constraints on the triangle logarithmic singularity of $\X\to D^0\bar{D}^{*0}\to\DDbar\pi^0$.

\section{Acknowledgement}
This work, based on data collected using the Belle detector, which was
operated until June 2010, was supported by 
the Ministry of Education, Culture, Sports, Science, and
Technology (MEXT) of Japan, the Japan Society for the 
Promotion of Science (JSPS), and the Tau-Lepton Physics 
Research Center of Nagoya University; 
the Australian Research Council including grants
DP180102629, 
DP170102389, 
DP170102204, 
DE220100462, 
DP150103061, 
FT130100303; 
Austrian Federal Ministry of Education, Science and Research (FWF) and
FWF Austrian Science Fund No.~P~31361-N36;
the National Natural Science Foundation of China under Contracts
No.~11675166,  
No.~11705209;  
No.~11975076;  
No.~12135005;  
No.~12175041;  
No.~12161141008; 
Key Research Program of Frontier Sciences, Chinese Academy of Sciences (CAS), Grant No.~QYZDJ-SSW-SLH011; 
Project ZR2022JQ02 supported by Shandong Provincial Natural Science Foundation;
the Ministry of Education, Youth and Sports of the Czech
Republic under Contract No.~LTT17020;
the Czech Science Foundation Grant No. 22-18469S;
Horizon 2020 ERC Advanced Grant No.~884719 and ERC Starting Grant No.~947006 ``InterLeptons'' (European Union);
the Carl Zeiss Foundation, the Deutsche Forschungsgemeinschaft, the
Excellence Cluster Universe, and the VolkswagenStiftung;
the Department of Atomic Energy (Project Identification No. RTI 4002) and the Department of Science and Technology of India; 
the Istituto Nazionale di Fisica Nucleare of Italy; 
National Research Foundation (NRF) of Korea Grant
Nos.~2016R1\-D1A1B\-02012900, 2018R1\-A2B\-3003643,
2018R1\-A6A1A\-06024970, RS\-2022\-00197659,
2019R1\-I1A3A\-01058933, 2021R1\-A6A1A\-03043957,
2021R1\-F1A\-1060423, 2021R1\-F1A\-1064008, 2022R1\-A2C\-1003993,
2019H1D3A1A01101787, and 2022R1A2B5B02001535;
Radiation Science Research Institute, Foreign Large-size Research Facility Application Supporting project, the Global Science Experimental Data Hub Center of the Korea Institute of Science and Technology Information and KREONET/GLORIAD;
the Polish Ministry of Science and Higher Education and 
the National Science Center;
the Ministry of Science and Higher Education of the Russian Federation, Agreement 14.W03.31.0026, 
and the HSE University Basic Research Program, Moscow; 
University of Tabuk research grants
S-1440-0321, S-0256-1438, and S-0280-1439 (Saudi Arabia);
the Slovenian Research Agency Grant Nos. J1-9124 and P1-0135;
Ikerbasque, Basque Foundation for Science, Spain;
the Swiss National Science Foundation; 
the Ministry of Education and the Ministry of Science and Technology of Taiwan;
and the United States Department of Energy and the National Science Foundation.
These acknowledgements are not to be interpreted as an endorsement of any
statement made by any of our institutes, funding agencies, governments, or
their representatives.
We thank the KEKB group for the excellent operation of the
accelerator; the KEK cryogenics group for the efficient
operation of the solenoid; and the KEK computer group and the Pacific Northwest National
Laboratory (PNNL) Environmental Molecular Sciences Laboratory (EMSL)
computing group for strong computing support; and the National
Institute of Informatics, and Science Information NETwork 6 (SINET6) for
valuable network support.


\begin{thebibliography}{**}
\bibitem{CCNote} Charge conjugate decay chains are included throughout this paper.
\end{thebibliography}
\end{document}